\documentclass{vgtc}                          % final (conference style)
\graphicspath{{figures/}{pictures/}{images/}{./}} % where to search for the images

\usepackage{times}                     % we use Times as the main font
\usepackage{tabu}                      % only used for the table example
\usepackage{booktabs}                  % only used for the table example
\usepackage{lipsum}                    % used to generate placeholder text
\usepackage{mwe}                       % used to generate placeholder figures

\usepackage{mathptmx}                  % use matching math font

\onlineid{0}

\vgtccategory{Research}

\vgtcinsertpkg

\title{Mental Well-being Opportunities in Interacting and Reflecting with Personal Data Sculptures of EEG}

\author{Maria Teresa Ortoleva\thanks{e-mail: maria-teresa.ortoleva@kcl.ac.uk}\\ %
        \scriptsize King's College London %
\and Rita Borgo\thanks{e-mail: rita.borgo@kcl.ac.uk}\\ %
     \scriptsize King's College London %
\and Alfie Abdul-Rahman\thanks{e-mail: alfie.abdulrahman@kcl.ac.uk}\\ %
     \parbox{1.4in}{\scriptsize \centering King's College London}}

\teaser{
    \centering
    \includegraphics[width=\linewidth]{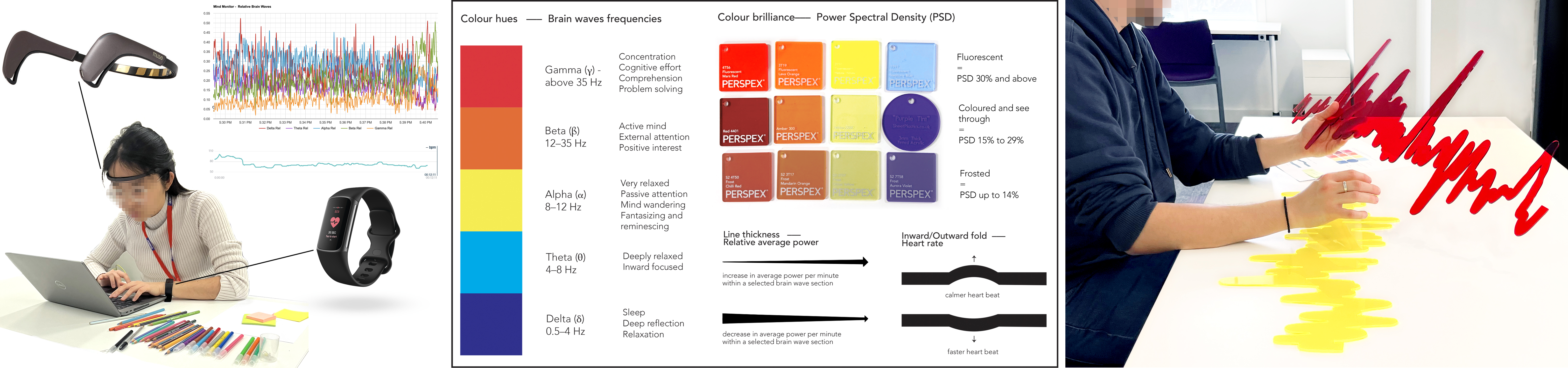}
    \caption{Study process: (a) Data collection of personal EEG, heart rate and narrative data; (b) Key to designing physical encodings for the fabrication of personal data sculptures; (c) Interaction with physicalized personal data, followed by reflective interview.}
    \label{fig:teaser:study-process}
}

\abstract{Data physicalization is a research area in quick expansion whose necessity and popularity are motivated by the pervasiveness of data in our everyday. While the reflective ability of personal data physicalization has been vastly documented, their mental health and emotional well-being benefits remain largely unexplored. We present a qualitative study where we create personal data sculptures of electroencephalograms (EEG) and mental activity, observe users' interactions with them, and analyze their reflections for hints of self-discovery and intended behavioral change. We argue that there is a ground for using personal data sculptures as prompts for reflection on mental well-being and motivators for self-caring, and that data sculptures for mental well-being are a finalized use of data physicalization worth exploring further.

} % end of abstract


\begin{document}


\firstsection{Introduction}

\maketitle

%% \section{Introduction} %for journal use above \firstsection{..} instead
Data physicalization has been the subject of increasing studies in recent years, establishing itself as a research area of its own~\cite{Jansen:2015:CHI, Dragicevic:2021:Springer}. In our \textit{`increasingly datafied world'}~\cite{Lupton:2017:Feeling-your-data}, physical manifestations of data – from everyday objects to displays and art~\cite{wiki:dataphys} – offer many \textit{`perceptual, cognitive, and societal benefits'}  through their unique ability to engage a public of any level of data literacy: procuring enticing sensory and inter-modal interactions, supporting meaningful cognitive explorations and learning, providing memorable experiences, sparking emotional response and reflection~\cite{Jansen:2015:CHI}. Within this area, personal data objects, and among them those produced with data from self-tracking devices and practices, especially offer an opportunity for many nuances in narrative and reflection~\cite{Karyda:2020:CHI:reflective-self-tracking, Karyda:2021:CHI:data-agents, Karyda:2021:IEEE-CGA:narrative}, even connecting people around the sharing of personal experience through data~\cite{Nissen:2015:data-things, Karyda:2021:CHI:data-agents}.

This context forms the premises on which, in this paper, we argue personal data physicalizations may potentially act as objects of well-being, supporting self-reflection and motivating self-care. Through their distinctive characteristics, physicalizations could play an active role in mental health-oriented conversations and interventions. However, the mental health and emotional well-being benefits of interacting and reflecting on one's personal data and with one's personal sculptures remain largely unexplored.

In this paper, we aim to test the ground for using personal physicalizations as tools for self-exploration and mental well-being. Our empirical study explores the potential of personal data sculptures in fostering self-discovery, generating insights and self-awareness about personal behavior, reflecting on time usage, and promoting self-care propositions. We worked closely with a small group of participants to collect, analyze, represent, and compare their EEG data from a high-concentration activity versus data from a moment of relaxation and wandering of the mind. Participants were presented with 3D physical or 2D visual representations of their mind activity data from those moments and invited to contemplate, compare, interact, and reflect on their effects in an interview. 

This study identified evidence that, through the very sensory, perceptual, cognitive, and reflective characteristics proper of the physicalization process, data sculptures could offer a promising tool for insight and reflection on one's own mental well-being and act as persuasive motivators to pursue activities that engage the mind in a positive and nourishing way, thus inspiring acts of self-care.

\section{Related Work}

\textbf{Narrative and reflection.}
 The variety of nuances of reflection afforded by data objects has been a focus in Karyda et al.~\cite{Karyda:2021:IEEE-CGA:narrative}. In their work, authors imagine and make personalized data crafts and data-encoded everyday objects and observe their agency% in the everyday
: from a modified foosball table to interactively reflect on personal data to modified ski boots of an affective value for reflective self-tracking~\cite{Karyda:2020:CHI:reflective-self-tracking}, to crafted gifts of personal biometric data for meaning-making and story-telling~\cite{Karyda:2021:CHI:data-agents}. The work reveals deep and nuanced reflective abilities that data physicalizations possess and may support an application within the scope of mental well-being.

\textbf{Self-tracking and personal physicalizations.} 
Other studies in data physicalization cross personal informatics and practices of self-tracking via wearable devices. Examples can be found in Stusak's Activity Sculptures transforming running data into trophy-like objects~\cite{Stusak:2014:activity-sculptures}, in Khot's SweatAtoms, combining multiple biometric data into laser-cut abstract shapes~\cite{Khot:2014:SweatAtoms}, and in Sauv\'{e}'s LOOP, animating and geometrically representing daily progress against a personal goal~\cite{Sauvé:2020:LOOP}. With a focus on physical health and activity rather than mental health, these examples explore the contribution to self-understanding and motivating a healthy lifestyle of accessing one's data in a physical form. 

\textbf{Tangible interfaces for mental health.} 
A separate strand in data physicalization is that of Tangible User Interfaces %, a recent development of which has been exploring
and their use for digital mental health interventions. 
An example is Farrall et al.'s investigation of Physical Artefacts for Well-being Support (PAWS)~\cite{Farrall:2023:breathing-ball}.  The authors propose the development of a biofeedback-based device that, while held by users, changes its shape in relation to their breathing, with applications in emotion control in a clinical setting. In approaching a mental well-being scope, personal data sculptures can learn from these neighboring device-based explorations and their applications.

\textbf{Previous work: Mindscapes.} 
The motivation for this study also comes from previous exploratory work as part of the Mindscapes project, an artist-academic collaboration supported by KCL Culture and King's Department of Informatics in 2020–2022. The project saw Ortoleva and Borgo collaborate to collect brainwaves of work/labor and leisure via a Muse 2~\cite{MUSE-2} portable EEG device and explore the physicalization of the signature patterns of the occurring mind states into sculpture and installation art~\cite{Culture:Mindscapes-residency}.

\section{Study Design}
We conduct an empirical study involving a small group of participants across two in-person sessions. We gather biometric and narrative data, create personalized data sculptures, and facilitate participant interaction with their data, followed by reflective semi-structured interviews (Fig.~\ref{fig:teaser:study-process}). The study was granted ethical clearance (MRPP-23/24-41174) by KCL's Research Ethics Committee.

\subsection{Participants' Sampling}
We targeted 18 to 24-year-old university students with no sensory impairment or currently experiencing symptoms of a mental health condition. The motivation behind the choice of age group is that – with `three-quarters of a lifetime mental health disorders starting by the mid-twenties'~\cite{Kessler:2007:age-of-onset}, and the transition to university life being a factor for increased sensitivity~\cite{Duffy:2019:Lancet} – they represent a focus group in the promotion of mental well-being awareness, prevention and early intervention. This target group resonated with the scope of our study and its potential future applications.

The study was advertised on King's College London's internal Fortnightly Circular: Research Volunteer Recruitment. From 77 applications, we selected a sample of 8 participants (P01–P08) to be split into two subgroups of 4: the first, labeled ``3D group'' (P01, P02, P04, and P06), would interact with personal data sculptures  and the second, labeled ``2D group'' (P03, P05, P07, and P08), would receive equivalent visualizations for control. Participants were selected to reflect a demographic sample from the university student population, and care was taken to balance the two subgroups.

\subsection{Session One: Data Collection}
We ran 60-minute individual data-collection sessions with participants. In the first part, the participants filled in a questionnaire rating how much they valued their mental well-being, what they considered important to it, and how much they considered themselves to know about the topic. They also completed a personal timetable, reflecting on how they split their weekly time between study occupations and leisure and relaxation. The aim was to build a baseline of their views and self-awareness at the start of the study. Participants wore portable EEG Muse headbands and Fitbit watches, used in the second part to record their heart rates (see Fig~\ref{fig:teaser:study-process}a). They were briefed on choosing a concentration task related to their coursework and a relaxation task meaningful to them, and they performed them for 12-18 minutes each. After each task, participants filled in a narrative timeline, marking memorable events, thoughts, gestures, or emotions and their perceived timing during the recording.

\begin{figure}
    \centering
    \includegraphics[width=1\linewidth]{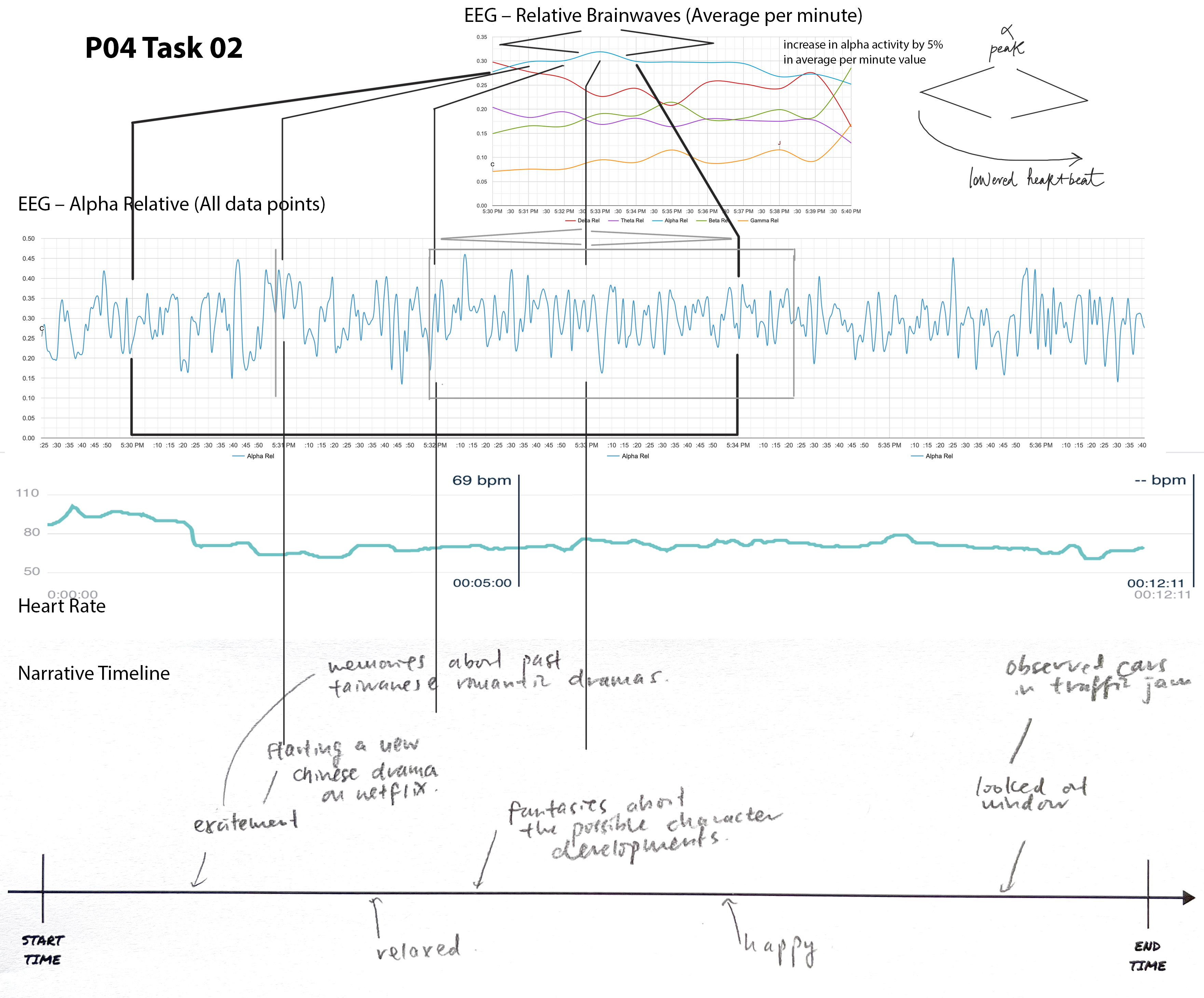}
    \caption{Alignment of EEG, heart rate, and narrative timeline}
    \label{fig:EEG-heart-timeline}
\end{figure}

\subsection {Data Sculptures Design and Prototyping}
With the collected EEG data, we used the Online Graphing feature of the Mind Monitor app~\cite{Clutterbuck:Mind-monitor}, plotting the five main waves (alpha, beta, gamma, delta, and theta) by relative Power Spectral Density (PSD) on the y-axis, and time on the x-axis. Along the time axis, we also aligned the heart rate graph from Fitbit data and the narrative timeline drawn by participants (see Fig~\ref{fig:EEG-heart-timeline}) - mindful that the latter is a subjective measure of perceived time. In order to present an amount of data that is significant, readable, and practical to fabricate into a sculpture, we selected segments of about 3 minutes across all the recordings. Within the EEG graph, the segment belonged to a single brainwave frequency from each reading whose power behavior (increase/decrease trends, peaks/troughs) represented what the participant was doing and their mental engagement. In the heart rate graph, we evaluated for any simultaneous variation to signify a calming trend or an intensifying effort. The subjective timeline provided valuable guidance in inferring notes on the state of mind, thoughts, and emotions the participant reportedly was experiencing around that time.

For the four participants in our 3D group, the selected EEG lines from their concentration and relaxation tasks were transformed into laser-cut Perspex shapes, see-through, and scaled to 70cm in width and 20–40cm in height, thus physically confronting the proportions of the body. The color of the Perspex would symbolize the frequency of the represented brainwave, and its relative power would translate into the degree of transparency and brightness of the material (frosted tones for low PSD to fluorescent tones for high PSD). The EEG line thickness was modulated to follow increase/decrease trends in relative power within the 3 minutes. Last, a relative increase/decrease in heartbeat was incorporated by heat-bending the Perspex shape outwards or inwards (extro/introflexion) (see Fig~\ref{fig:teaser:study-process}b). Similar encodings were designed for our 2D group to make similar visualizations and to allow \textit{`a fair alternative presentation method as a baseline for comparison'}~\cite{Jansen:2015:CHI}). In the 2D version, the waves were a digital graphic of the line, modulated according to the same parameters of color and thickness as the 3D, and printed on paper. The only differences were that the size was limited to the standard A3 printing format, and due to the lack of a third dimension, heart rate variation was represented by a cool/warm variation in the color temperature of the main wave color. Each data representation was labeled with a title extracted from the participants' narrative timelines, indicating a specific state of mind captured and made visible by each encoded line: \textit{e.g., ``In an effort to understand  (gamma),'' ``Connecting my thoughts (beta)'', ``Reminiscing'' (alpha), ``Inner speech while looking out the window (alpha)''}.

\begin{figure}
    \centering
    \includegraphics[width=1\linewidth]{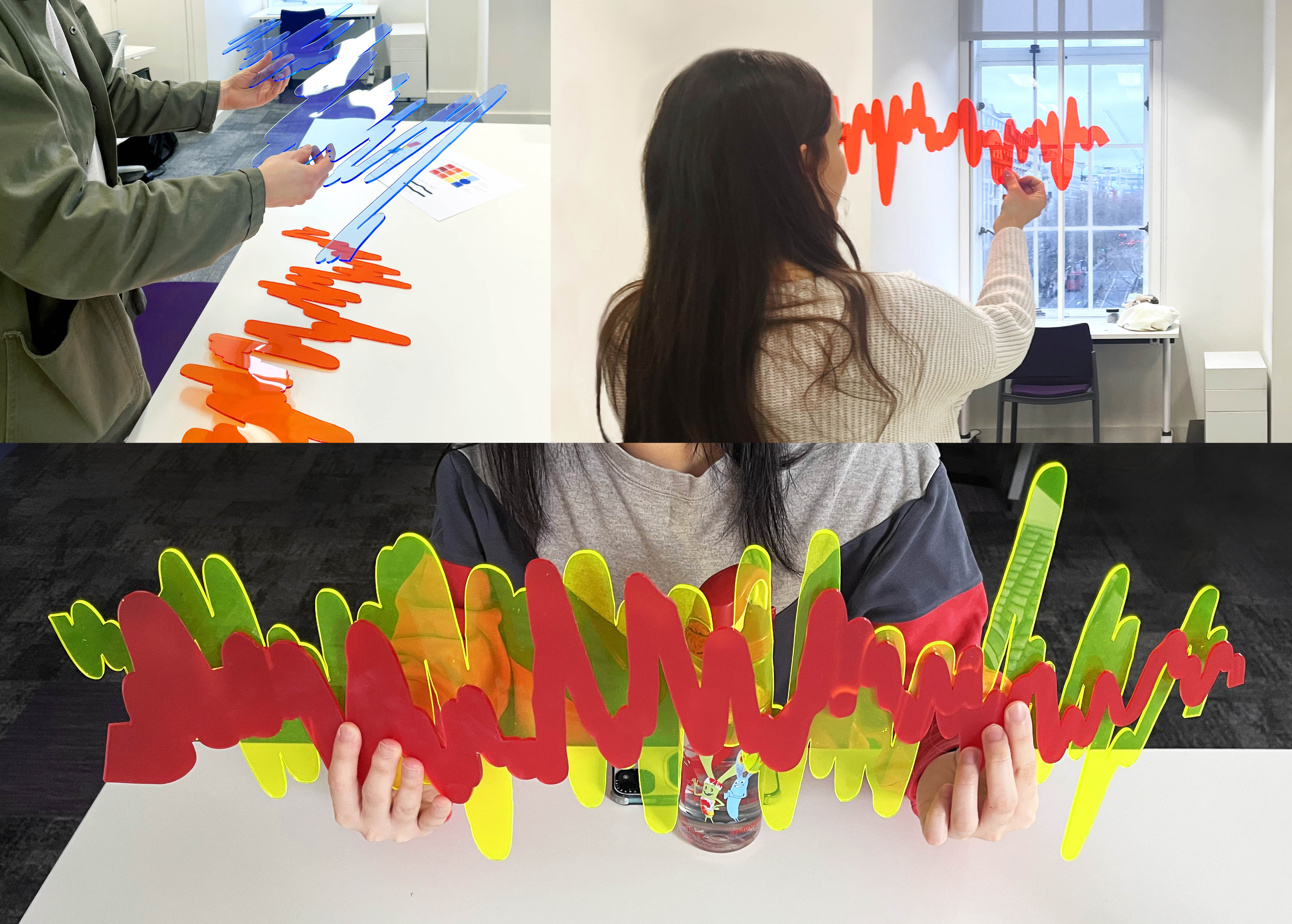}
    \caption{Gestures of interaction with personal data sculptures}
    \label{fig:interaction-shots}
\end{figure}

\subsection{Session Two: Interaction}
We re-invited participants to a second 60-minute individual session for interaction and reflection on their data representations. They were presented on a table with a key to the encodings, a seat, and ample space to move around them. Participants did 15 minutes of self-led interaction, exploring their two data representations from their concentration and relaxation activities.  Afterward, they engaged in a 30-minute semi-structured reflective interview. At intervals during the session, participants completed a Positive Affect Negative Affect Scale (PANAS)~\cite{Watson:1988:PANAS}, rating on a 5-point scale their emotions in the two weeks before the session (PANAS 1) immediately after interacting with their data (PANAS 2), and after completing the reflective interview (PANAS 3).

\subsection{Analysis}
From transcripts of the interviews and notes from direct observation of the interaction, we performed a thematic analysis of the participant's responses to encountering their data representations. The taxonomy for the analysis was based on evaluation criteria suggested by Jansen~\cite{Jansen:2015:CHI} and used in relatable studies~\cite{Karyda:2021:IEEE-CGA:narrative, Stusak:2014:activity-sculptures}, with some adaptations targeting reflection effects on mental well-being and behavior, as per our scope. We selected interviews for evidence of: ``Explorative actions, gestures, and senses'' (red), ``Feelings and emotions'' (blue), ``Ability to support recall'' (purple), ``Supporting cognition and self-discovery'' (yellow), ``Prompting reflection'' (turquoise), and ``Propositions of behavioral change'' (green). For full reference please see Supplementary Material.

PANAS scores after interaction (PANAS 2) and after reflection (PANAS 3) were compared to the initial baseline (PANAS 1). Thus, we focused on the percentage variation in positive and negative affect brought by the perceptual, cognitive, and reflective experience of encountering one's own data and the additional variation brought by articulating and verbalizing thoughts in the interview.

Overall, we had a double scope: (1) Seeing the effects of encountering personal data of one's mind activity as a physical sculpture over a visualization; (2) Tracing the benefits and opportunities for reflection on well-being and self-caring across the whole process of data collection, physicalization, interaction, and verbalization.

\section{Reflection}

\paragraph{Modalities of interaction} 
Reportedly, sight was the dominant sense used by both groups to explore their data; however, the accompanying attitude and gestures differed. The 2D group interaction was faster and more static. Their attitude and sitting postures suggested careful examination and minimal gestures. They concentrated extensively on the key, constantly relying on it to interpret their two representations. 3D group, instead, soon started pairing touch to sight: feeling the surface and flexion with their fingertips, holding the objects, raising them to eyes' height and towards the light of the window to appreciate the intensity of the colors (see Fig~\ref{fig:interaction-shots}). To perform these gestures, the participants stood up and used both arms and hands, thus engaging the whole body. Exploiting the material's transparency, they tried layering and comparing their two waves, accessing a more complex exploration than the 2D group. Their attention was absorbed mainly by the data sculptures, which offered a more intense opportunity for embodied cognition.

\paragraph{Emotional response} The most reported emotions in the interviews were interest and surprise. Data representations were received positively, with curiosity, and led to realize something new about the mind. However, the 2D group initially reported being \textit{`confused'} (P05) and \textit{`dumbfounded'} (P03), feelings that they overcome as they gradually understood their data, and only then felt interest and fascination kicking in (e.g. P05). The encounter was more immediately captivating for the 3D group, who enthusiastically explored their objects and used more emphatic language to report feeling \textit{`enjoyment'}, \textit{`fun'}, and \textit{`amusement'} (P01). We can speculate that the 2D group diligently overcame their initial distance because in a lab setting, however, they may have walked on if in a public context. Personal sculptures, instead, fascinated perception and emotions, seamlessly accompanying users into the harder task of cognitive exploration.

\paragraph{Supporting memory retrieval} %Similarly to Karyda's study with \textit{'Data Agents'}~\cite{Karyda:2021:CHI:data-agents}, 
We looked for evidence of the data representation acting as evocative triggers, and we asked distinctively for recall of facts and recall of emotions. All participants in both groups claimed that their data representation effectively supported reminiscing; however, participants in the 2D group made a more substantial effort to self-impress that they could retrieve events and emotions rationally but not feel them again (P05). Some doubted if the memory produced in the effort was a false memory. The 3D group was more fluent, confident, and detailed in recollecting. They pointed at waves' features and could make sense of them through specific feelings, thoughts, or events of their past experience. Memory retrieval led to feeling identification with their sculpture, and that their sculpture enabled them, in return, to dig deeper into their memory (P01).

\paragraph{Cognition and self-discovery}
Both groups found that the data representations aid in understanding and are informative (e.g. P08). They were amazed at their mind activity, saw that it challenged stereotypes, and – although the amount of information was necessarily limited – it sparked curiosity to learn more about the topic beyond the study (P02 and P04). Data offered feedback on personal activity: either revealing something new about oneself or strengthening self-impression into actual self-awareness. Most discoveries regarded the ability to focus intensely when tasked with it and the unexpected high power of alpha waves while the mind wanders and relaxes. However, the 2D group was slower to interpret the visualization and had a milder emotional impact (P08). 3D sculptures, instead, were seen to better intrigue (P04) and convey the complexity and dynamicity of the mind's activity. Some participants projected extra layers of meaning onto the shapes and colors, giving symbolic and metaphorical interpretations of their sculptures. In addition, their discovery process was invested with feelings of \textit{`gladness'}, \textit{`pride'} (P06), and \textit{`relief'} (P02). The 3D group described the information from their sculptures as concrete and tangible feedback, thus providing a solid starting point for reflection and change.

\paragraph{Self-identification and reflection}
Everyone felt described by their data representation, although only sculptures suggested an empathetic connection with one's data. Additionally, the data representations sparked reflective comments on performance and habits from everyone, but the 3D group experienced a more natural and intense flow of reflection. In contrast, the 2D group relied more on prompts from the interview questions. Once more, the concreteness of the 3D objects had a greater striking effect on people's imagination: \textit{`Seeing your mind activity in an object makes you reflect more, question yourself more'} (P04) and \textit{`I think I'm pretty self-reflective and self-aware about making sure I'm not anxious or making sure my mental well-being is good. But this has shown me physically how my brain is working, which is now making me think of ways to further improve my mental well-being. It is different to just think about it or to actually see it.'} (P06)

\paragraph{Propositions for change and effects on behavior}
Self-reflection and awareness naturally blended in with propositions for taking advantage of the eye-opening experience provided by the data to inform one's own future behavior. Asked to refer back to their initial timetable task and to propose changes, the participants described the desirable change as not the time distribution between study and relaxation but the pursuit of a true quality engagement of their mind in both areas (P06). Participants felt persuaded to pull away from their typical distractions to pursue deep focus (e.g. \textit{`Now that I have seen [their effect] actually in data, I have no excuse'} (P06)). They extensively reflected on how different occupations nurture positive mental engagement, leaving a visible trace in the EEG, and therefore are worth and necessary caring for. Many highlighted a difference between nourishing and compelling relaxation activities they performed in the study and lazy mind-numbing occupations they often slip into, such as phone scrolling (P01 and P03). Qualitatively, however, ideas for change across the 2D group were milder, less imaginative, and mostly referred to replicate the same performance they did in the study. Also, they envisaged barriers that would soon hinder their enthusiasm for change. By contrast, the 3D group was more visionary and articulate. Under the influence of their personal data sculptures, they felt driven and persuaded, and the importance of caring for their mental well-being felt more real. Some imagined scenarios where they would take their physicalization in their personal space to work as a visual reminder of self-care. \textit{`I would like to have my data sculptures displayed in my room as a reminder of my balance between studying and relaxing. I would use them as a starting point to work on my mental well-being.'} (P02) and \textit{`If I had this plastered on my wall in my room, then, whenever I look at it, it does make me think of,' `how is my brain feeling today?'. `It would help me reflect more.'} (P04)

%Despite the sculptures not providing a pathway to achieve it, it kindled curiosity and motivation to learn how to nurture it beyond the study's boundaries (e.g. P02).

\paragraph{Participant 07} P07's experience was different to the rest of the 2D group. She interacted with her data in the most object-like way afforded by printed sheets: holding them, moving them, and even layering them to bring interesting sections closer and help comparison. Although unaware of what the physicalizations for the 3D group looked like, she expressed regret that her visualizations were not transparent, as overlapping them would have helped her exploration. Interestingly, her attempt at a more physical interaction brought her experience closer to that described by the 3D group: she empathized more with her data, attached symbolic meanings to the color features, and imaginatively thought of how she could perform her studying and playing video games in \textit{`more conscientious ways'} that would procure positive mind engagement. The exceptionality of P07's interaction suggests that the embodied exploration procured by physical data has key benefits that offer a rich reflection on mental well-being data.

\begin{figure}
    %\centering    
%    \includegraphics[width=\linewidth]{Variation in PANAS scores after interaction and after verbal reflection_arrows-d.pdf}    
\includegraphics[width=\linewidth]{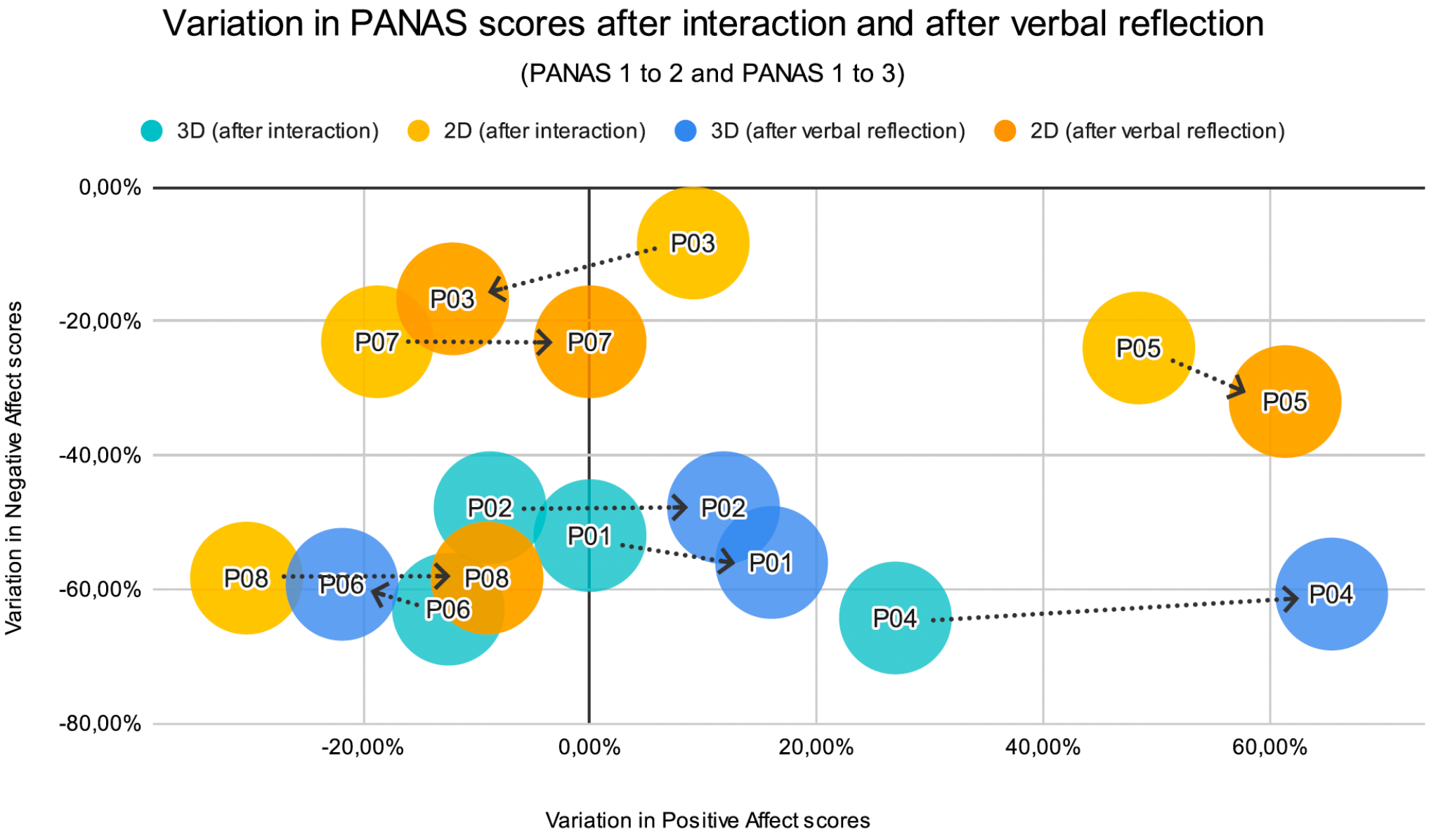}
    \caption{PANAS scores show a decrease in negative emotions upon encountering data representations and a milder increase for most participants in positive affect upon reflecting on their experience.}
    \label{fig:PANAS}
\end{figure}

\paragraph{PANAS} 
Although the benefits of positive affect vary between participants in the two groups, the experience of encountering and interacting with a representation of their data significantly decreased negative emotions. This drop is greater for people interacting with a data sculpture, with values lowering by 48–64 percent. Interviews kept negative emotions low, similarly to the interaction, but added a further increase in positive emotions for all participants except P03 and P06. We argue that verbalizing reflection facilitated clarity and self-awareness, potentially making most participants confident enough to choose higher affect scores. . Contextual to these data, we noted that although the 2D and 3D groups were demographically balanced, PANAS 1 profiled the 3D group as emotionally less positive and more negative than the 2D group. This note underscores the 3D group's shift from negative to positive emotions upon encountering and reflecting on their personal data sculpture, suggesting a calming effect. See Fig~\ref{fig:PANAS}.

%Other contextual note: coming into the session, students have reported feeling stressed because of the end of the term approaching. We see that the encounter with their data has provided a positive experience, and especially one of respite.

\section{Conclusion and Future Work}
The study shows positive hints that personal data sculptures for well-being are a tool worth exploring. Data objects trigger emotional and perceptual fascination, facilitating cognitive engagement with personal data, self-identification, and empathy. They depict mental well-being tangibly, encouraging reflection and motivating individuals to care for it persuasively. Furthermore, interaction with them has compelling and potentially grounding qualities.
%They show mental well-being as concrete, prompt reflection, and motivate to nurture self-caring behavior. 

These hints sustain our motivation to pursue future work in this area: to verify our initial findings, to assess their impact on longer-term behavior (via longitudinal studies), to test their ability to create a more compassionate culture around mental well-being, and perhaps even to offer a tool to contribute to pre-intervention strategies and non-clinical interventions in mental health.

Strands emerged while working on this study that could help further argue data objects for mental well-being: (1) the potential application of data sculptures in a public context, leveraging perceptual enticement to initiate positive conversations around mental health with a larger audience; (2) the potential to leverage data sculptures' sensory/perceptual playfulness to engage young people who, while more at risk, may also be reluctant to talk about their mental well-being; (3) the potential of data physicalizations to be used as material probes and prompts to facilitate reflection in communities and therapeutic settings.

\acknowledgments{
This work was supported by a training grant from the UK Research and Innovation: EPRSC DTP Studentship.}

\bibliographystyle{abbrv-doi}

\bibliography{template}
\end{document}